\documentclass[proof]{WileyASNA-v1}
\usepackage{framed}
\usepackage{natbib}
\usepackage{tablefootnote}

\usepackage{soul}
\usepackage{romannum}
\AtBeginDocument{\pagenumbering{arabic}}

\articletype{Original Article}

\received{Date}
\revised{Date}
\accepted{Date}

\begin{document}

\title{Modeling brightness temperature of sunspots using ALMA single-dish observations}

\author[1]{Filip Matkovi\'c*}

\author[1,2]{Roman Braj\v{s}a}

\author[3,4]{Arnold O. Benz}

\author[2]{Hans -G. Ludwig}

\author[5,6]{Caius L. Selhorst}

\author[7,8]{Ivica Skoki\'{c}}

\author[1]{Davor Sudar}

\author[9]{Arnold Hanslmeier}

\authormark{Matkovi\'{c} \textsc{et al}}

\address[1]{\orgdiv{Hvar Observatory}, \orgname{Faculty of Geodesy, University of Zagreb}, \orgaddress{\state{Zagreb}, \country{Croatia}}}

\address[2]{\orgdiv{Landessternwarte}, \orgname{Zentrum f\"ur Astronomie der Universit\"at Heidelberg}, \orgaddress{\state{Heidelberg}, \country{Germany}}}

\address[3]{\orgdiv{University of Applied Sciences and Arts Northwestern Switzerland}, \orgaddress{\state{Windisch}, \country{Switzerland}}}

\address[4]{\orgdiv{Institute for Particle Physics and Astrophysics}, \orgaddress{\state{ETH Zurich}, \country{Switzerland}}}

\address[5]{\orgdiv{NAT - N\'ucleo de Astrof\'isica}, \orgname{Universidade Cidade de S\~ao Paulo}, \orgaddress{\state{S{\~a}o Paulo}, \country{Brazil}}}

\address[6]{\orgdiv{Center for Solar-Terrestrial Research}, \orgname{New Jersey Institute of Technology, Newark}, \orgaddress{\state{New Jersey}, \country{USA}}}

\address[7]{\orgdiv{Astronomical Society} \orgname{"Anonymus"}, \orgaddress{\state{Valpovo}, \country{Croatia}}}

\address[8]{\orgdiv{Protostar Labs} \orgname{d.o.o.}, \orgaddress{\state{Beli{\v s}\'ce}, \country{Croatia}}}

\address[9]{\orgdiv{Institute of Physics}, \orgname{University of Graz}, \orgaddress{\state{Graz}, \country{Austria}}}

\corres{*Filip Matkovi\'{c}, Hvar Observatory, Faculty of Geodesy, University of Zagreb, Ka\v{c}i\'{c}eva 26, 10000 Zagreb, Croatia. \\\email{fmatkovic@geof.hr}}

\presentaddress{Croatian Science Foundation, project ID: 7549. Austrian-Croatian Bilateral Scientific Projects. Horizon 2020 project SOLARNET, project ID: 824135. European Union -- NextGenerationEU, National Recovery and Resilience Plan (NPOO). Alexander von Humboldt Foundation. S{\~a}o Paulo Research Foundation (FAPESP), grant number: 2019/03301-8.}

\fundingInfo{Croatian Science Foundation, project ID: 7549. Austrian-Croatian Bilateral Scientific Projects. Horizon 2020 project SOLARNET, project ID: 824135. European Union -- NextGenerationEU, National Recovery and Resilience Plan (NPOO). Alexander von Humboldt Foundation. S{\~a}o Paulo Research Foundation (FAPESP), grant number: 2019/03301-8.}

\jnlcitation{\cname{
\author{Matkovi\'c F.}, 
\author{Braj{\v s}a R.}, 
\author{Benz A. O.},
\author{Ludwig H.-G.},
\author{Selhorst C. L.},
\author{Skoki\'c I.},
\author{Sudar D.}, and 
\author{Hanslmeier A.}} (\cyear{Year}), 
\ctitle{Modeling brightness temperature of sunspots using ALMA single-dish observations}, \cjournal{Astronomische Nachrichten}, \cvol{Vol. No.}.}

\abstract{We model sunspot brightness temperature in the wavelength range 0.3 -- 10 mm using the Atacama Large Millimeter/submillimeter Array (ALMA) single-dish measurements together with other instruments. Our modeling idea starts from an existing model, the 1D semi-empirical Avrett-Tian-Landi-Curdt-W\"ulser (ATLCW) model for sunspots. The original model is then modified in density and temperature with a constant factor over all heights. A least-square fit in both density and temperature to the observed values yields an improved model for sunspot brightness temperature. Observations show that sunspots are darker than QS at wavelengths below 2 mm and brighter beyond, while fitting of the ATLCW sunspot model imply that sunspots should be darker than QS over the entire ALMA range. The best-fitted sunspot brightness temperature profiles resulted in 3.6 -- 4.0 times lower sunspot density compared to the original sunspot ATLCW model, or 1.4 -- 1.8 times higher density compared to QS. The temperature came out around 1.2 times higher than the original ATLCW predictions for sunspots, or 1.2 times lower than QS temperature. Even with improved density and temperature parameters, the remaining discrepancy between the best-fitted models and observations indicated that additional improvements are needed in the ATLCW model and modeling assumptions.}


\maketitle

\section{Introduction}
\label{Introduction}
Sunspots occur in medium- and large-sized active regions (ARs) and are recognized as compact regions with lower emission than the surrounding solar atmosphere when observed in visible light (e.g., Solanki \citeyear{Solanki2003}; Thomas \& Weiss \citeyear{Thomas2008}). Their darker appearance stems from a lower plasma temperature relative to their surroundings due to a strong magnetic field emerging from the solar interior and preventing hotter plasma from entering the sunspot region. In the photosphere, a typical sunspot consists of cool, dark umbra up to 2\;000 K cooler than the background and a magnetic field exceeding 1.5 kG, surrounded by warmer, brighter penumbra with a weaker magnetic field between 1 and 1.5 kG (e.g., de Oliveira e Silva et al. \citeyear{deOliveiraeSilva2022}). In the chromosphere and corona, the sunspot magnetic field weakens and forms bright loops with confined warm, dense plasma, giving ARs their characteristic 3D structure (e.g., Solanki \citeyear{Solanki2003}; van Driel-Gesztelyi \& Green \citeyear{vanDriel2015}).

Models that match the observed parameters help to understand the physics of sunspots. In this work, we consider semi-empirical solar atmosphere models, which usually specify plasma density and temperature profiles, constructed from various observed spectral lines, typically in the extreme-ultraviolet (EUV) and infrared spectrum. In the radio domain, such models have been widely applied to solar structures such as the quiet-Sun (QS), AR, coronal hole (CH), and less frequently prominence (PR) structures. Previous studies (e.g., Benz et al. \citeyear{Benz1997}; Selhorst et al. \citeyear{Selhorst2005}; Braj{\v s}a et al. \citeyear{Brajsa2007}, \citeyear{Brajsa2009}; Selhorst et al. \citeyear{Selhorst2017}, \citeyear{Selhorst2019}; Braj{\v s}a et al. \citeyear{Brajsa2018a}) employed semi-empirical models like the VAL (Vernazza et al. \citeyear{Vernazza1981}), FAL (Fontenla et al. \citeyear{Fontenla1993}), and SSC (Selhorst et al. \citeyear{Selhorst2005}) to model the intensity of radio emission and to infer plasma density and temperature within these solar structures.

Most previous sunspot modeling studies (e.g., Yun et al., \citeyear{Yun1984}; Maltby et al., \citeyear{Maltby1986}; Ding \& Fang, \citeyear{DingFang1989}; Socas-Navarro, \citeyear{SocasNavarro2007}; Fontenla et al., \citeyear{Fontenla2009}; Braj{\v s}a et al., \citeyear{Brajsa2018a}; umbral and penumbral models review in Loukitcheva \& Reardon, \citeyear{Loukitcheva2022}) were limited by insufficient observations or failed to reproduce the observed emission intensity. With the start of regular solar observations and the first solar science verification observations of the Sun by the Atacama Large Millimeter/submillimeter Array (ALMA) in late 2015, single-dish (White et al. \citeyear{White2017}) and interferometric observations (Shimojo et al. \citeyear{Shimojo2017}) with unprecedented spatial resolution across the 0.3 -- 10 mm wavelength range became available, significantly improving the observational basis for sunspot modeling.

The first semi-empirical modeling using ALMA data was performed by de Oliveira e Silva et al. (\citeyear{deOliveiraeSilva2022}), who took advantage of the new ALMA single-dish data to model the brightness temperature and plasma properties of facular, umbral, and penumbral regions of an AR (Table 2 in de Oliveira e Silva et al. \citeyear{deOliveiraeSilva2022}). They employed a genetic-algorithm based SSC modeling with ALMA full-disk observations at 100 and 230 GHz and the Nobeyama Radioheliograph (NoRH) full-disk observations at 17 GHz, achieving good agreement between calculated and observed sunspot brightness temperatures.

In our previous studies Matkovi\'c et al. (\citeyear{Matkovic2024}) and Matkovi\'c et al. (\citeyear{Matkovic2025}), hereafter Paper \Romannum{1} and Paper \Romannum{2}, respectively, we modeled brightness temperature across 0.3 -- 10 mm for the QS, AR devoid of sunspot (i.e., detected area of an AR with exclusion of any present sunspots), CH, and PR structures on the solar disk. The Avrett-Tian-Landi-Curdt-W\"ulser (ATLCW) QS model (Avrett et al. \citeyear{Avrett2015}) served as the reference solar atmosphere model, accurately reproducing QS brightness temperatures across the ALMA wavelength range and providing a solid basis for modeling other solar structures. Results from Papers \Romannum{1} and \Romannum{2} also confirm that the dominant radiation emission mechanism in the mm and sub-mm range is thermal bremsstrahlung, consistent with prior studies, and is assumed in the present work for modeling sunspot properties.

In this work, we extend brightness temperature modeling to sunspots using the semi-empirical ATLCW atmospheric model for the sunspot structure (Table 2 in Avrett et al. \citeyear{Avrett2015}). The ATLCW sunspot parameters served as input for our sunspot modeling. Section \ref{Calculation} gives a brief description of the modeling methodology using data and measurements obtained with ALMA and other radio instruments described in Section \ref{observational_results}. Section \ref{Results} presents a detailed description of the sunspot modeling results, while Section \ref{discussion_conclusion} further discusses these findings in comparison to previous studies and draws conclusions.

\section{Obtaining theoretical models and brightness temperature calculation}
\label{Calculation}
In this section, we give a brief overview of the procedure for modeling and calculating the sunspot brightness temperature. For more details see Papers \Romannum{1} and \Romannum{2}. 

To calculate the brightness temperature $T_\mathrm{b}$ at a given wavelength $\lambda$, we use the Rayleigh-Jeans approximation of Planck's law, which gives the observed emission intensity $I_\lambda$ proportional to the plasma temperature (e.g., Wilson et al. \citeyear{Wilson2013}). This leads to the following radiative transfer equation (e.g., Papers \Romannum{1} and \Romannum{2} and references therein):
\begin{equation}
\label{Eq_radiative_transfer}
T_\mathrm{b}(\lambda)=\int_{0}^{\infty}T_\mathrm{e}e^{-\tau_\lambda}{\rm d}\tau_\lambda
\end{equation}
Here, $T_\mathrm{b}(\lambda)=\lambda^4I_\lambda/(2ck_\mathrm{B})$, where $c$ and $k_\mathrm{B}$ are the speed of light and Boltzmann constant, respectively. The parameter $\tau_\lambda$ is the optical depth, which depends on the radiation emission mechanism at a given wavelength.

Previous works investigating the mm and sub-mm emission of the quiet and active chromosphere (Bogod \& Gelfreikh \citeyear{Bogod1980}; Dulk \citeyear{Dulk1985}; Zirin \citeyear{Zirin1988}; Hurford \citeyear{Hurford1992}; Braj{\v s}a \citeyear{Brajsa1993}; White \citeyear{White2002}; Benz \citeyear{Benz2009}; Wedemeyer et al. \citeyear{Wedemeyer2016}; Braj{\v s}a et al. \citeyear{Brajsa2018a}; Paper \Romannum{1}; Paper \Romannum{2}) suggest thermal bremsstrahlung as the dominant radiation mechanism in the 0.3 -- 10 mm range. Braj{\v s}a et al. (\citeyear{Brajsa2009}; \citeyear{Brajsa2018a}; and references therein) noted that within ARs, beside thermal bremsstrahlung, a thermal gyromagnetic contribution in the mm emission cannot be neglected if the magnetic field reaches kG range or higher. Such a strong magnetic field is unlikely at chromospheric heights observable by ALMA, even above sunspots. Consequently, gyromagnetic contributions are negligible and we assume thermal bremsstrahlung as the predominant sunspot radiation mechanism in the mm and sub-mm ranges. For a height increment d$h$, with $c=\nu\times\lambda$, where $\nu$ is the observed emission frequency, the optical depth d$\tau$ for thermal bremsstrahlung emission can then be expressed as (e.g., Benz \citeyear{Benz2002}):
\begin{equation}
\label{Eq_optical_depth}
{\rm d}\tau_\nu=\frac{0.01146\;\mathrm{cm^{5}Hz^{2}K^{3/2}}\times g \times n^{2}_{e}}{\left(1-8.06\times 10^{7}\;\mathrm{cm^3Hz^{2}}\times n_{e}/\nu^{2}\right)^{1/2}\nu^{2}T_{e}^{3/2}} {\rm d} h,
\end{equation}
where $n_\mathrm{e}$ and $T_\mathrm{e}$ are electron density and temperature, respectively, and $g$ is the Gaunt factor (Rybicki \& Lightman \citeyear{Rybicki1985}; Weinberg \citeyear{Weinberg2020}). The Gaunt factor is calculated using the interpolation method developed by van Hoof et al. (\citeyear{vanHoof2014}), later also implemented by Sim\~oes et al. (\citeyear{Simoes2017}) and Selhorst et al. (\citeyear{Selhorst2019}).

The electron density and temperature parameters as functions of height needed for calculating the brightness temperature are taken from the input atmospheric model. In this work, we use the semi-empirical ATLCW model as the input atmospheric model, which is available for two different solar regions: for the QS (Table 1 in Avrett et al. \citeyear{Avrett2015}) and sunspot regions (Table 2 in Avrett et al. \citeyear{Avrett2015}). We employ the latter model as input to our modeling of the brightness temperature and investigate the validity of the original ATLCW sunspot model of Avrett et al. (\citeyear{Avrett2015}) compared to actual observations.

It should be noted that the ATLCW models of Avrett et al. (\citeyear{Avrett2015}) are static non-local thermodynamic equilibrium (NLTE) models computed under statistical equilibrium and therefore neglect non-equilibrium (NEQ) ionization effects. Hydrogen ionization, and thus electron density, is assumed to respond instantaneously to local conditions, producing smooth, static electron density distributions that cannot capture the delayed or hysteretic behavior caused by shock dynamics (e.g., Carlsson \& Stein \citeyear{Carlsson2002}; Wedemeyer et al. \citeyear{Wedemeyer2016}). Although these models reproduce average electron densities and mean spectra well, they do not account for NEQ effects that can significantly influence mm-wave brightness, particularly in dynamic sunspot phenomena. Previous studies (e.g., Loukitcheva et al. \citeyear{Loukitcheva2014}; \citeyear{Loukitcheva2017}; da Silva Santos et al. \citeyear{daSilvaSantos2020}) have shown that NEQ effects can change mm brightness temperatures by several hundred to $\sim$1\;000 K in active sunspots. Since the sunspots analyzed here were relatively quiescent during the observing time period, NLTE and NEQ effects are not treated explicitly in our calculations. The implications of this approximation are discussed in Section \ref{discussion_conclusion}.

For a given wavelength $\lambda$, the brightness temperature is calculated using Equations \ref{Eq_radiative_transfer} and \ref{Eq_optical_depth}, with electron density and temperature for a given height $h$ taken from the ATLCW sunspot model. These height-dependent contributions to the brightness temperature are integrated over all heights considered for the respective ATLCW sunspot model, specifically between 0 and 58\;100 km. The upper height boundary corresponds to the upper cut-off height of the ATLCW sunspot model (Table 2 in Avrett et al. \citeyear{Avrett2015}), above which the contribution of plasma properties to the chromospheric emission is negligible. To increase accuracy of the brightness temperature calculation, the original ATLCW sunspot model parameters were interpolated by a factor of 100 across the considered height range. This procedure is applied to all observed wavelengths, producing a brightness temperature profile, which characterizes sunspot emission from 0.3 to 10 mm.

The ATLCW model was modified only by applying scaling factors to the density and temperature. While the original model assumes hydrostatic equilibrium, it does not include magnetohydrostatic effects, and our adjustments likewise ignore contributions from the magnetohydrostatic balance. This simplification is justified because our study focuses on deriving the vertical stratification of the brightness temperature from observed average sunspot properties. In this context, magnetic pressure and tension effects can be safely ignored, as we do not explicitly couple the gas and magnetic pressures through a force-balance condition. The emphasis is on radiative transfer calculations (Equation \ref{Eq_radiative_transfer}) rather than reproducing the underlying mechanical equilibrium.

To derive a brightness temperature profile that best fits the actual measurements, we apply the $\chi^2$-minimization technique (e.g., Ivezi\'c et al. \citeyear{Ivezic2014}) to find the global minimum of a sum of $\chi^2$ values across all observed wavelengths. This is done by varying the input electron density and temperature parameters using multiplicative factors (Papers \Romannum{1} and \Romannum{2}):
\begin{equation}
\label{multi_factors}
f_\mathrm{n}=\frac{n_\mathrm{e}\mathrm{(new)}}{n_\mathrm{e}\mathrm{(original)}},\; f_\mathrm{T}=\frac{T_\mathrm{e}\mathrm{(new)}}{T_\mathrm{e}\mathrm{(original)}},
\end{equation}
where $n_\mathrm{e}$ and $T_\mathrm{e}$ denote electron densities and temperatures of the original and modified (new) ATLCW models. The same multiplicative factor was applied uniformly to all heights during sunspot modeling. The result is the best-fitted brightness temperature profile with the corresponding density and temperature scaling factors that provide the best estimate of the average sunspot density and temperature relative to the original ATLCW sunspot model. We do not apply height-dependent scaling factors, as doing so would introduce an excessive number of free parameters, which we want to avoid.

\section{Observational results}
\label{observational_results}

\subsection{ALMA data}
\label{ALMA_data}
We use single-dish full-disk solar observations obtained in the ALMA 0.3 -- 10 mm range. The primary data are from the ALMA Bands 3 ($\lambda=2.80$ mm) and 6 ($\lambda=1.21$ mm). Interferometric observations are not used due to their limited availability, field-of-view and time constraints, and insufficient data points for reliable model fitting. Single-dish data instead provide a larger, more consistent sample with broadly comparable spatial resolution across instruments, and mixing them with interferometric data is avoided because of their fundamentally different spatial resolutions.

The sunspot is extracted from image using a method similar to that for AR extraction in Paper \Romannum{1}. Since the area of an AR, not including sunspots, is bright in ALMA images, it was extracted by detecting pixels with a brightness temperature above a specified threshold. The corresponding sunspot, being darker than the surrounding AR area, is automatically excluded from the rest of the AR area, and its boundaries are detected. Namely, the brightness temperature of such isolated sunspot region is then determined by averaging the brightness temperature of all pixels within its detected boundary, that is, below the threshold used for extraction of the surrounding AR area.

With the extraction procedure used here, the measured sunspot brightness temperature represents the combined emission from the umbra and penumbra. These components are treated jointly to maintain consistency with previous evaluations of 1D ATLCW models (Papers \Romannum{1} and \Romannum{2}), which assessed the ability of atmospheric models to reproduce the global properties of solar structures. The sunspot modeling in Section \ref{Results} follows the same methodology. In addition, mm single-dish observations are strongly affected by convolution-induced spatial smearing, which mixes emission from neighboring regions. This effect is particularly pronounced for the penumbra, making it difficult to distinguish from the umbra, especially in ALMA Band 3 images where sunspots often appear predominantly umbral. As a result, to separate penumbral measurements and modeling would be unreliable, whereas umbral properties are more robust. For these reasons, our approach differs from some previous sunspot studies (e.g., Severino et al. \citeyear{Severino1994}; Loukitcheva et al. \citeyear{Loukitcheva2017}; Loukitcheva \& Reardon \citeyear{Loukitcheva2022}; de Oliveira e Silva \citeyear{deOliveiraeSilva2022}; and references therein).

For ALMA Band 6, which also encompasses both umbra and penumbra, we adopt the sunspot brightness temperature measured by Braj{\v s}a et al. (\citeyear{Brajsa2018b}). In addition, we analyze the ALMA Band 3 full-disk observations ($\lambda=2.80$ mm, beam size = 58") of the same sunspot as Braj{\v s}a et al. (\citeyear{Brajsa2018b}), observed one day earlier on December 17, 2015. The corresponding measured brightness temperatures of a sunspot in ALMA Bands 3 and 6 are given in Table \ref{Table_data}.

To compensate for the limb brightening effect (e.g., Sudar et al. \citeyear{Sudar2019}), we first the brightness temperature of a QS region in the center of the solar disk far from the solar limb by averaging the brightness temperature within a 15-pixel radius of that QS region. We then measure the brightness temperature of a QS region of 10-pixel radius at the same radial distance from the solar center as the sunspot and compare it to the measurement of the given sunspot. The QS brightness temperatures at a similar radial distance as the sunspot and at the solar disk center are denoted with $T_\mathrm{b}$(QS) and $T_\mathrm{b}$(QS$_\mathrm{c}$), respectively, in Table \ref{Table_data}. The difference between the sunspot and the non-central QS brightness temperature is given by $\Delta T_\mathrm{b}=T_\mathrm{b}$(sunspot) $-$ $T_\mathrm{b}$(QS) in the same table. Adding $\Delta T_\mathrm{b}$ to $T_\mathrm{b}$(QS$_\mathrm{c}$) yields the sunspot brightness temperature corrected for the limb brightening effect.
\begin{table*}[h!]
\centering
\caption{~Sunspot brightness temperatures $T_\mathrm{b}$(sunspot) measured by ALMA, JCMT, and Nobeyama at given times (Date) and wavelength/frequency ($\lambda / \nu$), and spatial resolutions (Beam size). The brightness temperature $T_\mathrm{b}$(QS) corresponds to the brightness temperature measurement of a QS region at a similar radial distance from the solar disk centre as the sunspot, while $\Delta T_\mathrm{b}=T_\mathrm{b}$(sunspot) $-$ $T_\mathrm{b}$(QS). $\Delta T_\mathrm{b}$ is used to obtain $T_\mathrm{b}^\mathrm{a}$(sunspot) and $T_\mathrm{b}^\mathrm{b}$(sunspot) for measurement procedures "a" and "b" (see Section \ref{Results} for details), which are given in bold in this table and plotted in Figure \ref{SS_ATLCWSS}, and are used for fitting the calculated brightness temperature from the ATLCW sunspot model. For details on individual observations see corresponding references (Reference).}
\label{Table_data} 
\centering
\resizebox{2\columnwidth}{!}{
\begin{tabular}{c c c c c c c c c c c}
\hline\midrule
Instrument & Date & $\lambda$ / $\nu$ & Beam size & $T_\mathrm{b}$(QS) ($T_\mathrm{b}$(QS$_\mathrm{c}$)) & $T_\mathrm{b}$(sunspot) & $\Delta T_\mathrm{b}$ &$\boldsymbol{T_\mathrm{b}^\mathrm{a}}$\textbf{(sunspot)}&$\boldsymbol{T_\mathrm{b}^\mathrm{b}}$\textbf{(sunspot)}& Reference \\
(name) & (y:m:d) & (mm) / (GHz) & (arcsec) & (K) & (K) &  (K) &\textbf{(K)}&\textbf{(K)}& (citation)\\
\midrule
JCMT    &  1991-05-08   &  0.35 / 857 & 17.6 & 4\;790& 3\;780 &  $-1\;010$       &$\boldsymbol{=T_\mathrm{b}}$\textbf{(sunspot)}&\textbf{3\;910}&  Lindsey \& Kopp (\citeyear{Lindsey1995}) \\ 
\midrule
JCMT    &  1991-05-08   &  0.85 / 353 & 18.4 & 6\;000& 5\;200 &  $-800$       &$\boldsymbol{=T_\mathrm{b}}$\textbf{(sunspot)}&\textbf{4\;870}&  Lindsey \& Kopp (\citeyear{Lindsey1995}) \\ 
\midrule
ALMA    &  2015-12-18   &  1.21 / 248 & 26 & 6\;170 (6\;040)& 6\;080 &  $-90$       &\textbf{5\;950}&\textbf{6\;110}&  Braj{\v s}a et al. (\citeyear{Brajsa2018b}) \\ 
\midrule
Nobeyama  &  2014-02-12   &  2.60 / 115 & 15 & 7\;440 & 7\;460 &  $+20$  &$\boldsymbol{=T_\mathrm{b}}$\textbf{(sunspot)}&\textbf{6\;930}& Iwai et al. (\citeyear{Iwai2016})   \\
\midrule
ALMA    &  2015-12-17    &  2.80 / 107 & 58 & 7\;240 (7\;110)& 7\;280 &  $+40$       &\textbf{7\;150}&\textbf{7\;010}&   This work \\ 
\midrule
Nobeyama  &  2014-02-12   &  3.50 / 85 & 19 & 7\;900 & 7\;830 &  $-70$ &$\boldsymbol{=T_\mathrm{b}}$\textbf{(sunspot)}&\textbf{7\;080}& Iwai \& Shimojo (\citeyear{Iwai2015})   \\
\midrule
Nobeyama  &  2014-02-12   & 8.80 / 34 & 10 & 8\;300 & 9\;300 &  $+1\;000$  &$\boldsymbol{=T_\mathrm{b}}$\textbf{(sunspot)}&\textbf{9\;160}& Iwai et al. (\citeyear{Iwai2016})   \\
\midrule
\end{tabular}
}
\end{table*}

\subsection{Other radio instruments}
Most of the currently available solar ALMA data cover only two short wavelengths within the ALMA 0.3 -- 10 mm range, leaving most of this range, especially longer mm wavelengths, unexplored. To gain a better understanding of the sunspot behavior in the ALMA wavelength range, we combine ALMA data with full-disk measurements from other radio telescopes.

For wavelengths shorter than ALMA Bands 3 and 6, we use sunspot measurements by Lindsey \& Kopp (\citeyear{Lindsey1995}), which are based on sub-mm observations obtained with the 15-m James Clerk Maxwell Telescope (JCMT)\footnote{\url{https://www.eaobservatory.org/jcmt/}}at the Mauna Kea Observatory in Hawaii, USA. We selected observations from May 8, 1991, which provide brightness temperature measurements of a sunspot seen both at 0.35 mm (beam size = 17.6") and 0.85 mm (beam size = 18.4"). As these measurements separately report umbral and penumbral average brightness temperatures, we combine them together by taking the simple average of their measured average brightness temperatures to ensure consistency with the ALMA data.

To extend coverage to longer mm wavelengths, we use measurements from the Nobeyama Radio Observatory\footnote{\url{https://www.nro.nao.ac.jp/en/}}. Specifically, we adopt sunspot brightness temperature from Iwai \& Shimojo (\citeyear{Iwai2015}), obtained with the Nobeyama 45-m parabolic aperture radio telescope on February 12, 2014 at 3.50 mm (beam size = 19"). The sunspot was detected within a narrow equatorial belt encompassing the target AR region (Figure 1 in Iwai \& Shimojo (\citeyear{Iwai2015})). Unlike the ALMA and JCMT data, this measurement represents the average of the brightness temperature over the umbral region only, as no penumbral brightness temperature measurements were available.

Although combining umbral and penumbral emission at some wavelengths but not others may seem inconsistent, the Nobeyama umbral measurement is included to constrain the long mm-wavelength end of the ALMA range, where models are otherwise poorly bounded and potentially unstable. Long mm-wavelength data are therefore valuable, though the resulting models must be interpreted with care given the mixed observational inputs. Beam smearing can further reduce penumbral contributions by mixing them with umbral and surrounding emission, whereas umbral emission is largely preserved and dominates single-dish observations. As a result, adding the Nobeyama umbral measurement does not significantly affect the inferred sunspot brightness temperature, and the resulting models primarily reflect umbral conditions.

Additional Nobeyama measurements were obtained with the Nobeyama Radioheliograph (NoRH), which consists of 84 parabolic antennas, each 80 cm in diameter, mounted on 490 m long lines in the east-west direction and 220 m long lines in the north-south direction. We use mm-wavelength sunspot observations from Iwai et al. (\citeyear{Iwai2016}), obtained from a full-disk solar image taken on February 12, 2014, just like Iwai \& Shimojo (\citeyear{Iwai2015}), but at 2.60 mm (beam size = 15") and 8.80 mm (beam size = 10") wavelengths. As in Iwai \& Shimojo (\citeyear{Iwai2015}), these measurements represent the brightness temperature averages over the umbral region only. All Nobeyama measurements are summarized in Table \ref{Table_data}. Together with the ALMA and JCMT observations, the Nobeyama measurements provide sufficient coverage of the ALMA 0.3 -- 10 mm range used in this study.

Unlike the earlier ALMA data, the JCMT and Nobeyama measurements are already corrected for the limb brightening effect. Accordingly, in Table \ref{Table_data} we have $T_\mathrm{b}$(QS) = $T_\mathrm{b}$(QS$_\mathrm{c}$), and $\Delta T_\mathrm{b}$ directly represents the brightness temperature difference between sunspot and QS regions as if both were located at the solar disk center, away from limb effects.

We should note that the $\chi^2$-minimization fitting accounts for measurement uncertainties. Previous studies (e.g., Shimojo et al. \citeyear{Shimojo2017}) estimated ALMA single-dish errors at $\sim5-10$\%, while more recent works (e.g., Alissandrakis et al. \citeyear{Alissandrakis2022}, \citeyear{Alissandrakis2023}) report a wider range ($0.3-10.6$\%). To remain consistent with both earlier and recent estimates, we adopt a uniform 5\% uncertainty, corresponding to the lower bound of earlier estimates, for all sunspot brightness temperature measurements. Although uncertainties generally increase with wavelength due to reduced spatial resolution, we apply the same 5\% error to all data, not only the ALMA measurements, given the overall increase of brightness temperature with wavelength.

\section{Modeling results and comparison with observations}
\label{Results}

\begin{figure}[h!]
\centering
\resizebox{0.99\hsize}{!}{\includegraphics{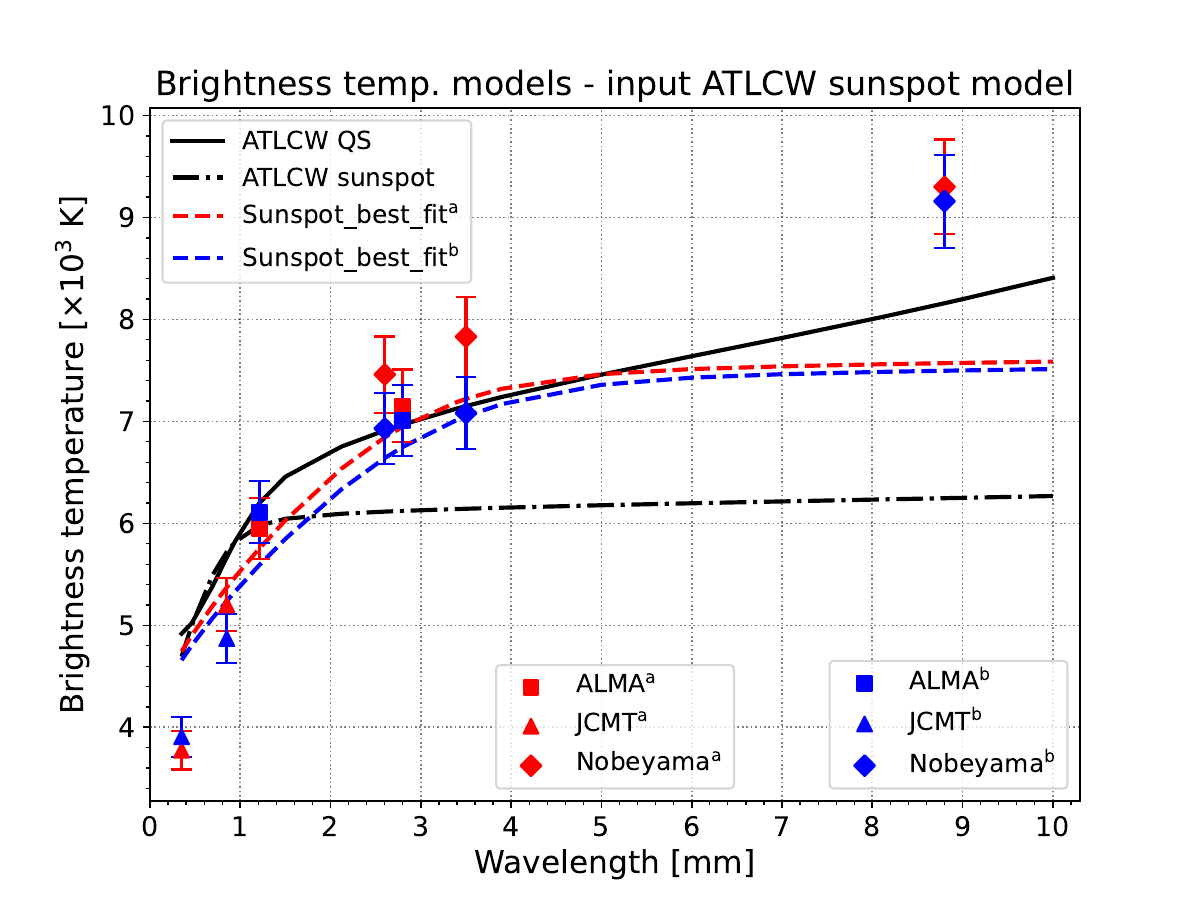}}
\caption{~Calculated sunspot brightness temperature profiles obtained from the original ATLCW sunspot model (black dash-dotted curve) and modified ATLCW sunspot model (dashed Sunspot$\_$best$\_$fit curves) fitted to plotted ALMA, JCMT, and Nobeyama sunspot brightness temperature measurements obtained using measurement procedures "a" (red symbols) and "b" (blue symbols). Error bars for all measurements correspond to the assumed 5\% measurement error for radio observations. The profile for the original ATLCW QS model (black solid curve) was adopted from Paper \Romannum{1}.}
\label{SS_ATLCWSS}
\end{figure}

We use two sunspot measurement sets for modeling. The first consists of actual measurements obtained by adding $\Delta T_\mathrm{b}$ from Table \ref{Table_data} to the central QS brightness temperature ($T_\mathrm{b}$(QS$_\mathrm{c}$)). These measurements and later models are shown with red symbols and the superscript "a". The second set uses the same $\Delta T_\mathrm{b}$ added to the QS brightness temperature given by the brightness temperature profile obtained for the original, unmodified ATLCW QS model (solid black curve in Figure \ref{SS_ATLCWSS}) for a given wavelength. We call this second set differential measurements and represent them and later models with blue symbols and the superscript "b". This approach tests the sensitivity of the modeling to small measurement differences. Below, we give the modeling results based on the fitting of the modified ATLCW sunspot models to these sunspot measurements.

Figure \ref{SS_ATLCWSS} presents sunspot modeling results based on the semi-empirical ATLCW sunspot model. At wavelengths shorter than 2 mm, the measurements of the sunspot brightness temperature are significantly lower than those of the QS (solid black brightness temperature profile in Figure \ref{SS_ATLCWSS} obtained from the ATLCW QS model), with the largest difference at the shortest wavelengths, corresponding to the chromospheric heights near the photosphere. In contrast, at wavelengths longer than 2 mm, the sunspot becomes increasingly brighter than the QS, with the largest differences occuring at the longest wavelengths, corresponding to heights near the transition region and the corona.

Despite error-bar uncertainties, sunspot brightness temperature generally increases with wavelength. Below 2 mm, measurements show clear stratification with small uncertainties, rising sharply from 3\;600 K to 6\;200 K over 0.86 mm and remaining mostly below the QS level. Between 2 and 4 mm, larger and overlapping error bars reduce the apparent stratification, especially for differential measurements, which show little variation ($6\;600-7\;500$ K) and an unclear contrast with the QS. In contrast, actual measurements exhibit a clearer increase ($6\;800-8\;200$ K), indicating brightness above the QS. The 8.8 mm measurement is distinctly higher ($8\;600-9\;800$ K) than shorter wavelengths, though the small difference relative to 3.5 mm ($\sim400$ K) suggests that the increase in brightness stratification flattens at longer mm wavelengths.

This result is consistent with our expectations, as well as other published studies (e.g., Loukitcheva et al. \citeyear{Loukitcheva2017}) indicating a similar behavior, where sunspots appear darker than the QS at photospheric and lower chromospheric heights, but brighter at greater chromospheric and coronal heights. Accordingly, the sunspot is darker at shorter ALMA wavelengths and brighter at longer ALMA wavelengths.

Comparison of the calculated best-fitted brightness temperature profiles and observations (Figure \ref{SS_ATLCWSS}) reveals a substantial discrepancy. The sunspot brightness temperature profile obtained from the original, unmodified ATLCW sunspot model (solid dash-dotted curve in Figure \ref{SS_ATLCWSS}) reproduces well the measured data points (red and blue symbols in Figure \ref{SS_ATLCWSS}) only near 1 mm wavelength. Generally, at wavelengths below 2 mm, it predicts a sunspot brightness temperature comparable to or lower than the QS given by the ATLCW QS model (black solid curve in Figure \ref{SS_ATLCWSS}), while at longer wavelengths it deviates strongly from all measurements, yielding much lower sunspot brightness temperatures and only a weak increase to about 6\;000 K, far below the observed values.

Although the original ATLCW sunspot model poorly reproduces the observed sunspot brightness temperatures, we explored whether modifying the original ATLCW sunspot model in electron density and temperature could improve the fit. The resulting best-fitted sunspot brightness temperature profiles obtained by modifying the ATLCW sunspot model show substantially better agreement with sunspot measurements for both measurement procedures "a" (red dotted curve in Figure \ref{SS_ATLCWSS}) and "b" (blue dotted curve in Figure \ref{SS_ATLCWSS}), though some discrepancies still remain.

At wavelengths below 2 mm, the best-fitted sunspot profiles agree well with sunspot measurements for both procedures "a" and "b", except for the JCMT 0.35 mm data, which are significantly lower. Taking into account the error bars, between 2 and 4 mm, most measurements, especially the differential measurements (blue symbols in Figure \ref{SS_ATLCWSS}), align closely with the corresponding best-fitted sunspot brightness temperature profiles. In this range, the best-fitted sunspot brightness temperature profile for the actual observations (red dotted curve in Figure \ref{SS_ATLCWSS}) only slightly exceeds the QS level beyond 3 mm.

Beyond 5 mm, all best-fitted sunspot profiles increasingly deviate from the expected behavior. The modeled sunspot profiles show a slow increase in brightness temperature but remain below the QS level. This discrepancy is most pronounced at 8.8 mm, where the observed sunspot brightness temperature is far higher than the prediction of the best-fitted sunspot model. Thus, at these wavelengths, the modeled sunspot brightness temperature profiles incorrectly suggest that the sunspot region is darker than the QS.
\begin{table*}[h!]
\centering
\caption{~Output multiplicative scaling factors for electron density ($f_n$ best fit) and temperature ($f_T$ best fit) corresponding to the best-fitted sunspot model for two measurement procedures ("a" and "b"). The corresponding minimum $\chi^2_\mathrm{min}$ value determined from the $\chi^2$-minimization method is also provided. The $f_n$ range and $f_T$ range correspond to the input density and temperature factor ranges (Equation \ref{multi_factors}) over which the density and temperature parameters of the input atmospheric model were varied to determine $\chi^2_\mathrm{min}$. A detailed description of the fitting procedure and scaling factor calculation is provided in Section \ref{Calculation}.}
\label{Table_2} 
\centering
\resizebox{2\columnwidth}{!}{
\begin{tabular}{c c c c c | c c c}
\hline
\multicolumn{5}{c|}{Input parameters}&\multicolumn{3}{c}{Output parameters}\\
\midrule
Input atmospheric model & Height range & $f_n$ range & $f_T$ range &Procedure& $f_n$ best fit & $f_T$ best fit  & $\chi^2_{\mathrm{min.}}$\\
(name) & (km) & (number) & (number) & (index)& (number) &(number)  &(number)\\
\hline\midrule
\multirow{2.5}{*}{ATLCW sunspot}    &  \multirow{2.5}{*}{$0-58\;100$}  & \multirow{2.5}{*}{$0.1-3$}&\multirow{2.5}{*}{$0.1-3$}&a& $0.28^{+0.03}_{-0.03}$ & $1.24^{+0.03}_{-0.03}$& $46.56$\\[0.5em]
&&&  &b&  $0.25^{+0.03}_{-0.03}$ &  $1.23^{+0.02}_{-0.03}$& $34.36$ \\
\midrule
\end{tabular}
}
\end{table*}

The best-fitted sunspot profiles yielded multiplicative scaling factors (Table \ref{Table_2}), with density factors $f^{a,b}_n$ best fit indicating that the average chromospheric sunspot plasma density is $3.5-4$ times lower than the prediction of the original, unmodified ATLCW sunspot model. Similarly, the temperature factors $f^{a,b}_T$ best fit indicate that the average chromospheric sunspot plasma temperature is about $23-24$\% higher than predicted by the original ATLCW sunspot model. This highlights a substantial discrepancy between the best-fitted sunspot brightness temperature profile and predictions of the original ATLCW sunspot model, especially in density.

The best-fitted multiplicative scaling factors compare the modified ATLCW sunspot model only to the original sunspot model, not to the QS. Using chromospheric (200 -- 1\;300 km; estimated for sunspot by de Oliveira e Silva et al. \citeyear{deOliveiraeSilva2022}) averages of the electron density and temperature for the original ATLCW QS and sunspot models (Tables 1 and 2 in Avrett et al. \citeyear{Avrett2015}), the modified sunspot model yields $1.40-1.56$ times higher average sunspot density and $1.19-1.20$ times higher average sunspot temperature compared to the QS. This indicates that the chromospheric sunspot is generally denser and hotter than the surrounding QS, possibly influenced by its strong magnetic field.

\section{Discussion and conclusions}
\label{discussion_conclusion}

The interpretation presented in this section is based on a simplified modeling framework that employs static, 1D atmospheric models computed under statistical equilibrium and magnetohydrostatic assumptions. While this approach allows us to isolate and interpret average properties and large-scale trends in the observed mm and sub-mm emission, it necessarily neglects time-dependent, multi-dimensional, and NEQ processes known to be present in the solar atmosphere. In particular, the use of static NLTE atmospheric models, as discussed in Section \ref{Calculation}, neglects time-dependent hydrogen ionisation and assumes magnetohydrostatic equilibrium. Past studies (e.g., Carlsson \& Stein \citeyear{Carlsson2002}; Leenaarts et al. \citeyear{Leenaarts2007}; Golding et al. \citeyear{Golding2016}) have demonstrated that NEQ ionisation can play an important role even in QS regions and substantially affect mm brightness temperatures through its influence on electron densities. Although the sunspots studied here were relatively quiescent and did not show pronounced dynamic signatures during the observing time period, NEQ effects are unlikely to be entirely negligible and may introduce systematic uncertainties in the absolute brightness temperatures inferred in this work. We therefore emphasize that the conclusions drawn here are most robust in a relative sense and for large-scale trends, while the absolute values should be interpreted with appropriate caution. Future studies incorporating time-dependent magnetohydrodynamic models with self-consistent NLTE/NEQ ionisation will be required to fully quantify these effects.

In a summary, the measured sunspot was darker below 2 mm, but brighter for longer wavelengths than the QS level defined by the ATLCW QS model (Figure \ref{SS_ATLCWSS}). Considering the error-bar uncertainties, the contrast in the brightness temperature between the sunspot and the ATLCW QS level firstly decreases toward 2 mm, remained near or slightly above the QS up to 3.5 mm, and rised toward 8.8 mm. This indicates a transition from a darker ($\lambda<2$ mm) to a brighter ($\lambda>2$ mm) sunspot region relative to the QS. Near the photosphere, strong sunspot magnetic field supresses emission, while near and at coronal heights, the sunspot expands into bright loops with a weaker magnetic field. This demonstrates that the reversal of the sunspot radiation emission relative to the QS occurs in the chromosphere at short mm wavelengths.

In the 2 -- 4 mm range, the measured sunspot brightness temperatures exhibit considerable variability around the QS level, with no clear trend with wavelength. Considering the error-bar uncertainties, the actual ALMA measurement (red square in Figure \ref{SS_ATLCWSS}) at 2.8 mm lies below nearby Nobeyama measurements (red diamonds in Figure \ref{SS_ATLCWSS}) at 2.6 and 3.5 mm, aligning more closely with the QS level. The uncertainties of the differential measurements (blue symbols in Figure \ref{SS_ATLCWSS}) show minimal deviation from the QS, with the 3.5 mm Nobeyama sunspot slightly darker than the QS level, likely because the 45-m Nobeyama radio telescope sampled only a portion rather than the entire umbra, let alone the entire sunspot region.

For actual measurements, the Nobeyama measurements at 2.6 and 3.5 mm include only umbral emission, which likely explains the brighter appearance than the ALMA measurement at 2.8 mm, which lies between. If the entire umbra and penumbra had been included, as in other measurements in this study, this slight reduction of the differential sunspot brightness temperature relative to the QS would probably be smaller, bringing the Nobeyama measurement closer to the QS level. 

The apparent differences may also reflect the higher spatial resolution of the Nobeyama radio telescope, which enhances contrast and reveals greater emission variability within a region, as commonly seen in interferometric observations. In contrast, single-dish measurements with larger beam sizes smooth out such variations, making them nearly or entirely undetectable. The beam sizes of all measurements used in this study exceed 10", which reduces any intrinsic emission variability below the measurement error bars, as the sunspot measurements average the emission over the entire sunspot. Thus, the reported error bars effectively encompass potential fluctuations in sunspot emission.

In this work, we observed the entire sunspot, combining both umbra and penumbra, to determine its average brightness temperature. This may partly explain the notable deviation between the modeled brightness temperature and Nobeyama 8.8 mm measurement in Figure \ref{SS_ATLCWSS}, which represents only umbral emission. The Nobeyama spatial resolution at 8.8 mm is sufficient to resolve sunspot structure, and including the penumbral contribution could reduce this discrepancy, bringing the measurement closer to the overall sunspot average.

Loukitcheva et al. (\citeyear{Loukitcheva2017}) analyzed the same AR as in our work, but using ALMA interferometric mosaics, resolving umbra and penumbra. They found the umbra $\sim$700 K cooler than the QS at 1.3 mm and $\sim$600 K brighter at 3 mm wavelength, while the penumbra exhibited opposite but weaker contrasts. Consequently, the average sunspot radiation emission is moderated in both directions (lower or higher brightness temperature), though umbral emission dominates. Our ALMA measurements closely resemble the umbral behavior reported by Loukitcheva et al. (\citeyear{Loukitcheva2017}), with lower contrast likely due to mixing of umbra and penumbra and differences in spatial resolution. These results suggest that single-dish observations primarily reflect umbral characteristics of the sunspot.

A more detailed examination of the sunspot substructure, as in Loukitcheva et al. (\citeyear{Loukitcheva2017}), would require ALMA interferometric observations to better resolve umbra and penumbra, allowing each substructure to be individually analyzed and modeled. This level of analysis is beyond the scope of the present study. Even with inclusion of interferometric ALMA observations, limitations persist. Although more ALMA bands are offered today for solar observations, not all ALMA bands observe the same sunspot, and observations are often non-simultaneous. Most ALMA solar observations to date focus on Bands 3 and 6, sometimes separated by one or two days. Ideally, a single sunspot observed across all bands would improve our analysis, but this is currently unfeasible due to the limited number of solar targets and available observations.

Future studies will aim to extend the wavelength coverage, improving the existing ALMA bands and adding new ones to probe solar heights from 400 to over 1\;700 km (Figure 5 in Wedemeyer et al. \citeyear{Wedemeyer2016}). New and improved projects for single-dish solar observations are also planned, such as the 50-m single-dish radio telescope Atacama Large Aperture Submillimeter Telescope (AtLAST) (e.g, Booth et al. \citeyear{Booth2024}; Wedemeyer et al. \citeyear{Wedemeyer2024}, \citeyear{Wedemeyer2025}). AtLAST will not reach the ALMA interferometric resolution, but will provide higher spatial and temporal resolution than the current single-dish ALMA observations, while compensating for broader wavelength coverage.

The profile for the original ATLCW sunspot model diverges markedly from the measurements starting around 2 mm, underestimating brightness temperatures at longer mm wavelengths. The best-fitted profiles of the modified ATLCW sunspot model show improved agreement up to $\sim$4 mm, but they still substantially underestimate the long-mm emission. These discrepancies indicate that both original and modified ATLCW sunspot models do not fully capture the physical properties governing sunspot brightness temperatures at long mm wavelengths.

Although the $\chi^2_\mathrm{min.}$ results (Table \ref{Table_2}) suggest that measurement procedure "b" is slightly preferred over "a", both yield similar sunspot density and temperature scaling factors. This highlights the sensitivity of the sunspot brightness temperature profile to variations in the density and temperature parameters of the atmospheric ATLCW sunspot model. Overall, redefining these physical parameters and further refining the ATLCW model are necessary to reproduce realistic sunspot brightness temperatures across the mm range.

Several issues may affect the sunspot models (i.e., brightness temperature profiles) obtained in this study. First issue could lie in the possibly simplified sunspot modeling approach, which applied the same multiplicative scaling factor to density and temperature individually over all heights simultaneously. Comparing with the SSC model parameters for the QS and sunspot regions, as provided in Table 2 of de Oliveira e Silva et al. (\citeyear{deOliveiraeSilva2022}), shows that density and temperature ratios between the QS and sunspot vary with height. One aspect of improving our current sunspot models could therefore require height-dependent scaling factors, which will be explored in future work to determine the most appropriate approach.

Another concern is the strong sunspot magnetic field. Equations \ref{Eq_radiative_transfer} and \ref{Eq_optical_depth} assume purely thermal bremsstrahlung emission and neglect magnetic effects. However, as noted in previous studies (e.g., Braj{\v s}a et al. \citeyear{Brajsa2009}; \citeyear{Brajsa2018a}; \citeyear{Brajsa2018b}), a strong and variable magnetic field can modify plasma density and temperature in sunspots, thereby affecting the optical depth and the resulting bremsstrahlung emission. Including magnetic field effects could therefore help reduce discrepancies between calculated models and observations.

A strong magnetic field may also enhance gyroresonance radiation in sunspots, which is neglected in our calculations. Previous studies like de Oliveira e Silva et al. (\citeyear{deOliveiraeSilva2022}) show that including gyroresonance emission improves agreement between modeled and observed sunspot brightness temperatures, while models based solely on thermal bremsstrahlung can underestimate emission and even predict umbrae darker than the QS at long wavelengths. A similar underestimation is present in our models at longer mm wavelengths. Other studies (e.g., Braj{\v s}a et al. \citeyear{Brajsa2009}; Rom{\v s}tajn et al \citeyear{Romstajn2009}; Braj{\v s}a et al. \citeyear{Brajsa2018a}) indicate that thermal gyroresonant emission becomes significant only for kG-level magnetic fields. For the ALMA-observed sunspot, the photospheric umbral field reaches $2.4-2.6$ kG, while the penumbra ranges from 0.4 to 1.1 kG, suggesting that a small gyroresonant contribution is possible. However, the magnetic field decreases with height, limiting its influence on mm emission. For ALMA Bands 3 and 6, non-negligible gyroresonant emission within the first five gyroresonant harmonics would require magnetic fields far exceeding typical solar values, making such contributions negligible at these wavelengths. At longer mm wavelengths, such as 8.8 mm observed by Nobeyama, the required magnetic field strength falls within the measured photospheric range from the fifth gyroresonant harmonic onward, allowing for a modest gyroresonant contribution. Future work will incorporate this component to quantify its effect on the modeled brightness temperatures and assess its role in matching observations.

The study by de Oliveira e Silva et al. (\citeyear{deOliveiraeSilva2022}) report that, in the chromosphere, umbra and penumbra have on average 24\% and 39\% higher density and 3\% and 5\% higher temperature, respectively, compared to the QS. These SSC model results are in a good quantitative and qualitative agreement with our ATLCW modeling results of the sunspot as a whole, with the derived densities within the SSC ranges, while the temperatures lie slightly outside the values reported for individual umbral and penumbral components of the SSC model. Like the SSC models, our sunspot models are based on single-dish observations and thus complement the work by de Oliveira e Silva et al. (\citeyear{deOliveiraeSilva2022}) by providing additional constraints toward a more realistic description of the mm sunspot emission.

In the low solar atmosphere, sunspots are small, compact, and effectively 2D structures, that are difficult to resolve with single-dish observations. In the absence of interferometric data, high-resolution UV and EUV imaging from the Interface Region Imaging Spectrograph (IRIS) and SDO can help define sunspot boundaries. These boundaries can be overlaid as contours on mm-wavelength images to measure sunspot emission within more physically reliable regions.

Regardless of the resolving ability, 1D sunspot models, such as the ATLCW model, have inherent limitations. Loukitcheva \& Reardon (\citeyear{Loukitcheva2022}) provide a concise overview of earlier 1D sunspot models. Among the models shown in Figure 3 of that study, only the model of Severino et al. (\citeyear{Severino1994}) demonstrated a good agreement with ALMA Bands 3 and 6 and with the results of de Oliveira e Silva et al. (\citeyear{deOliveiraeSilva2022}). While most 1D models reviewed in Loukitcheva \& Reardon (\citeyear{Loukitcheva2022}) reproduce the qualitative trend of sunspots transitioning from darker at short wavelengths to brighter at longer mm wavelengths relative to the QS, they fail to quantitatively match observed brightness temperatures across the 0.3 -- 10 mm range. Similar applies to the ATLCW sunspot model.

Figure 3 in Loukitcheva \& Reardon (\citeyear{Loukitcheva2022}) primarily address umbral measurements and models, with penumbral data and models shown separately in Figure 4 of that study. However, the existing 1D penumbral models do not adequately reproduce the measurements and differ substantially in their thermal profiles. Our results indicate that single-dish measurements, which combine umbral and penumbral emission, more closely follow umbral model behavior. This is consistent with comparisons to ALMA interferometric results reported by Loukitcheva et al. (\citeyear{Loukitcheva2017}) and reflects convolution effects at mm wavelengths, which suppress penumbral contributions and leave umbral emission as the dominant signal.

Like earlier 1D models, the ATLCW modeling adopts only a single number for brightness temperature at each height point, averaged over the entire sunspot, and thus does not capture small-scale emission variability seen in high-resolution observations. This averaging is adequate when variability is rapid, spatially scattered, or low-amplitude, as the integrated emission over the sunspot region remains stable. However, 1D modeling may be less reliable when variability is slow, large-amplitude, and localized, since averaging can obscure physically important features by mixing quiet and active parts. If such variability is present across all wavelengths, averaged profiles remain informative, otherwise, the physical fidelity of the model may be reduced. This limitation is most pronounced when comparing different sunspots or epochs, but can be mitigated by selecting sunspots observed during comparable quiet or active activity states, as done in this study.

It is fair to say that most, if not all, 1D models struggle to reproduce chromospheric conditions above sunspots, largely due to the limited number of suitable diagnostics, their complex formation processes, and the difficulty of inferring physical properties. The original ATLCW sunspot model used here is subject to these limitations. Although parameter scaling can alleviate some discrepancies, significant deviations from observations remain. Overall, the ATLCW sunspot model of Avrett et al. (\citeyear{Avrett2015}) is insufficient to accurately describe sunspots and requires fundamental modifications. Such modifications could improve the physical representation of sunspots, analogous to the successful representation of the QS region by the ATLCW QS model, as confirmed in Paper \Romannum{1}. As a result, the present conclusions are most robust for relative comparisons and wavelength-dependent trends, whereas the absolute values of the derived physical parameters and their applicability to individual dynamic sunspot features remain limited by the adopted assumptions.

Once needed improvements are implemented in the sunspot modeling, a natural next step is to extend the current 1D to 2D semi-empirical sunspot modeling. In a 2D framework, brightness temperature would be modeled for individual sunspot elements (e.g., image pixels), yielding spatially resolved 2D maps of brightness temperature and plasma density and temperature. This also enables separate analysis of umbra and penumbra. Applying this procedure across multiple wavelengths would further allow extension to 3D modeling, producing height-dependent stacks of 2D maps and revealing the vertical stratification sunspot properties. Extending ALMA observations to longer wavelengths would further probe higher atmospheric layers and provide tighter constraints on sunspot physical parameters.

\section*{Acknowledgments}

This work was supported by the Croatian Science Foundation as part of the "Young Researchers' Career Development Project - Training New Doctoral Students" under the project 7549 "Millimeter and submillimeter observations of the solar chromosphere with ALMA". Support from the Austrian-Croatian Bilateral Scientific Projects ”Comparison of ALMA observations with MHD-simulations of coronal waves interacting with coronal holes”, ”Multi-Wavelength Analysis of Solar Rotation Profile”, and "Analysis of solar eruptive phenomena from cradle to grave" is also acknowledged. It has also received funding from the Horizon 2020 project SOLARNET (824135, 2019–2023). This work was also supported by the European Union -- NextGenerationEU within the framework of the National Recovery and Resilience Plan (NPOO), project “Eruptive processes on the Sun”. In this paper, ALMA data ADS/JAO.ALMA\#2011.0.00020.SV were used. ALMA is a partnership of ESO (representing its member states), NSF (USA) and NINS (Japan), together with NRC (Canada), MOST and ASIAA (Taiwan), and KASI (Republic of Korea), in cooperation with the Republic of Chile. The Joint ALMA Observatory is operated by ESO, AUI/NRAO, and NAOJ. We thank the ALMA project for enabling solar observations with ALMA. RB acknowledges financial support from the Alexander von Humboldt Foundation. CLS acknowledges financial support from the S{\~a}o Paulo Research Foundation (FAPESP), grant number 2019/03301-8.

\appendix
\bibliography{Matkovic_et_al_sunspot_brightness_temp_model_final}

\end{document}